\documentclass[prd,reprint,aps,amsmath,amssymb,showpacs,nofootinbib]{revtex4-1}
\usepackage{graphicx,hyperref,color}
\newcommand{\bra}[1]{\langle #1|}
\newcommand{\ket}[1]{|#1 \rangle}

\newcommand{\Br}{{\mathrm{Br}}}

\begin{document}

\title{A Renormalizable Model for the Galactic Center Gamma Ray Excess from Dark Matter Annihilation}

\author{Seyda~Ipek}
\affiliation{Department of Physics, University of Washington, Seattle, WA 98195, USA}

\author{David~McKeen}
\affiliation{Department of Physics, University of Washington, Seattle, WA 98195, USA}

\author{Ann~E.~Nelson}
\affiliation{Department of Physics, University of Washington, Seattle, WA 98195, USA}

\begin{abstract}
Evidence for an excess of gamma rays with ${\cal O}(\rm GeV)$ energy coming from the center of our galaxy has been steadily accumulating over the past several years. Recent studies of the excess in data from the Fermi telescope have cast doubt on an explanation for the excess arising from unknown astrophysical sources. A potential source of the excess is the annihilation of dark matter into standard model final states, giving rise to gamma ray production. The spectrum of the excess is well fit by 30~GeV dark matter annihilating into a pair of $b$ quarks with a cross section of the same order of magnitude as expected for a thermal relic. Simple models that can lead to this annihilation channel for dark matter are in strong tension with null results from direct detection experiments. We construct a renormalizable model where dark matter-standard model interactions are mediated by a pseudoscalar that mixes with the CP-odd component of a pair of Higgs doublets, allowing for the gamma ray excess to be explained while suppressing the direct detection signal. We consider implications for this scenario from Higgs decays, rare $B$ meson decays and monojet searches and also comment on some difficulties that any dark matter model explaining the gamma ray excess via direct annihilation into quarks will encounter.
\end{abstract}

\maketitle

\section{Introduction}
\label{sec:intro}
One of the prime unanswered questions about our Universe is the nature of dark 
matter (DM). Evidence for DM is overwhelming, coming from a diverse set 
of observations, among them galactic rotation curves, cluster merging, and the 
cosmic microwave background (see, e.g.~\cite{Feng:2010gw} and references therein). So far, dark matter has not been observed 
nongravitationally, 
yet the fact that a thermal relic with weak-scale annihilation cross section into 
standard model (SM) final states would have an energy density today that is 
compatible with dark matter measurements offers hope that this will be possible. 
The nongravitational interactions of DM are being searched for at particle 
colliders, in direct detection experiments, and in so-called indirect detection 
experiments, where the products of DM annihilation or decay are sought.

One final state in particular that is searched for in indirect detection 
experiments is gamma rays which can be produced by DM annihilating, 
either (i) directly to photons, which would result in an unambiguous line in the case 
of two body decays, or (ii) into other SM particles that then decay and produce 
photons in the cascade. The Fermi collaboration has published limits on DM annihilation  \cite{Ackermann:2012qk,*Ackermann:2013uma,*Ackermann:2013yva,*Murgia} into final states containing photons.

Recently, evidence for such a signal has been mounting, with several 
groups~\cite{Daylan:2014rsa,Hooper:2013nhl,Hooper:2010mq,Goodenough:2009gk,*Hooper:2011ti,Abazajian:2012pn,*Gordon:2013vta,*Abazajian:2014fta} 
analyzing data from 
the Fermi Gamma Ray Space Telescope and finding an excess of gamma rays of energy $\sim$1-3~GeV in the region of the 
Galactic Center. This excess has a spectrum and spatial morphology 
compatible with DM annihilation. The potential for astrophysical backgrounds, in 
particular millisecond pulsars in this case, to fake a signal 
is always a worry in indirect 
detection experiments. However, the observation that this gamma ray excess 
extends quite far beyond the Galactic Center lessens the possibility of 
astrophysical fakes~\cite{Hooper:2013nhl}, with recent studies finding the excess to extend to at least $10^\circ$ from the Galactic Center~\cite{Daylan:2014rsa,Hooper:2013rwa,*Huang:2013pda}.

The excess's spectrum has been fit by DM annihilating 
to a number of final states, depending 
on its mass, notably 10~GeV DM annihilating to $\tau^+\tau^-$ (and possibly other leptons) \cite{Barger:2010mc,*Hardy:2014dea,*Modak:2013jya,Hooper:2010mq,Abazajian:2012pn,*Gordon:2013vta,*Abazajian:2014fta,Marshall:2011mm,*Buckley:2011mm,*Lacroix:2014eea} 
and 30~GeV DM to $b\bar b$ \cite{Goodenough:2009gk,*Hooper:2011ti,Barger:2010mc,*Hardy:2014dea,*Modak:2013jya,Abazajian:2012pn,*Gordon:2013vta,*Abazajian:2014fta,Berlin:2014tja,*Agrawal:2014una}. The size of the excess is compatible 
with an annihilation cross section roughly equal to that expected for a thermal 
relic, $\langle \sigma v_{\rm rel}\rangle=3\times10^{-26}~{\rm cm^3}/{\rm s}$, suggesting that it is actually the result of DM annihilation.

30~GeV DM that annihilates to $b$ quarks is particularly interesting, 
primarily because direct detection experiments have their maximal sensitivity 
to spin-independent interactions between nuclei and DM at that mass. 
Reconciling the extremely strong limit from direct detection in this mass range, 
presently $8\times10^{-46}~{\rm cm}^2$~\cite{Akerib:2013tjd}, with a potential indirect detection signal poses a challenge, possibly 
offering a clue about the structure of the SM-DM interactions. We will focus 
on this DM mass and final state in this paper.

In the case of DM annihilation to SM fermions through an 
$s$-channel mediator, 
we can roughly distinguish the distribution of final states by the spin of the 
mediator. Spin-0 mediators tend to couple with strength proportional to mass---either 
due to inheriting their couplings from the Higgs or because of general considerations 
of minimal flavor violation---which results in decays primarily to the heaviest fermion 
pair kinematically allowed. On the other hand, spin-1 mediators generally couple more 
democratically, leading to a more uniform mixture of final states.
For this reason, the fact that the excess is well fit by 30~GeV DM annihilating 
dominantly to $b\bar b$ suggests annihilation through a scalar. However, this is 
problematic: to get an appreciable indirect detection signal today requires scalar 
DM (fermionic DM annihilating through a scalar is $p$-wave suppressed) but 
this leads to a spin-independent direct detection 
cross section that is in conflict with experimental bounds, as mentioned above.
Therefore, we are led to consider a pseudoscalar mediator, instead of a scalar, 
between the (fermionic) DM and the SM, 
leading to an effective dimension-six operator of the form
\begin{align}
{\cal L}_{\rm eff}&=\frac{m_b}{\Lambda^3}\bar\chi i\gamma^5\chi \bar b i\gamma^5 b,
\label{eq:Leff}
\end{align}
where $\chi$ is the DM. This  
operator has been singled out previously as a good candidate to describe the 
effective interaction between the SM and the dark sector~\cite{Boehm:2014hva,Alves:2014yha}. 
It implies $s$-wave DM
annihilation, which allows the gamma ray excess to be fit while having a 
large enough suppression scale $\Lambda$ that it is not immediately ruled out 
by collider measurements of monojets/photons. The direct detection signal from this 
operator  is spin-dependent and velocity-suppressed, rendering it safe 
from current constraints.

To move beyond the effective, higher dimensional operator in Eq.~(\ref{eq:Leff}) 
requires confronting electroweak symmetry breaking because the SM 
portion of ${\cal L}_{\rm eff}$ is not an electroweak singlet:
\begin{align}
\bar b i\gamma^5 b=i\left(\bar b_L b_R-\bar b_R b_L\right).
\end{align}
Therefore, ${\cal L}_{\rm eff}$ has to include the Higgs field (which would make 
it a singlet) which then gets a vacuum expectation value (VEV), implying a 
mediator which can couple to the Higgs.

It is easy to construct a scalar-scalar interaction between DM and the SM 
using the ``Higgs portal" 
operator $H^\dagger H$, where $H$ is the SM Higgs doublet, since it is a SM 
gauge singlet. This portal 
has been well explored in the literature, particularly in its connection to DM~\cite{Burgess:2000yq,*Strassler:2006im,*Patt:2006fw,*Strassler:2006ri,*Pospelov:2011yp,*Bai:2012nv,*Walker:2013hka}. In this paper, however, we expand the 
Higgs sector of the SM to include a second doublet, 
which has enough degrees of freedom to allow for a pseudoscalar to mix with the dark 
matter mediator. In the presence of CP violation one could also induce a pseudoscalar-scalar coupling via this portal, however it is puzzling why a new boson with CP violating couplings would not also have a scalar coupling to the dark fermion. Including two Higgs doublets allows CP to be an approximate symmetry of the theory, broken by the SM fermion Yukawa coupling matrices. Tiny CP violating couplings will need to be included  in order to renormalize the theory at high orders in perturbation theory, but we simply assume that all flavor and CP violation is derived from  spurions proportional to the Yukawa coupling matrices, and so has minimal effect on the Higgs potential  and dark sector. 

The outline of this paper is as follows. In Sec.~\ref{sec:model} we introduce the two Higgs doublet model (2HDM) and the pseudoscalar mediator which mixes with the Higgs sector. We also discuss CP violation in the dark sector and in interactions between DM and SM fermions. We briefly discuss the annihilation cross section for our DM model in Sec.~\ref{sec:ann}. In Sec.~\ref{sec:results}, we catalog constraints on this model, such as direct detection, Higgs and $B$ meson decays, and monojets. Section~\ref{sec:conc} contains our conclusions.

\section{The Model}
\label{sec:model}
\subsection{CP-Conserving Extended Higgs Sector}
As mentioned above, a straightforward way to couple dark matter to the SM through pseudoscalar
exchange is by mixing the mediator with the pseudoscalar Higgs in a 2HDM.

For concreteness, we take the DM to be a Dirac fermion, $\chi$, with mass $m_\chi$, coupled to a real, gauge
  singlet, 
pseudoscalar  mediator, $a_0$, through
\begin{align}
{\cal L}_{\rm dark}&=y_\chi a_0 \bar\chi i\gamma^5\chi.
\label{eq:dm-yuk-flavor}
\end{align}
The mediator couples to the SM via the Higgs portal in the scalar potential which is
\begin{align}
&V=V_{\rm 2HDM}+\frac{1}{2}m_{a_0}^2a_0^2+\frac{\lambda_a}{4}a_0^4+V_{\rm port},
\\
&V_{\rm port}=iBa_0H_1^\dagger H_2+{\rm h.c.}
\label{eq:Vport}
\end{align}
with $H_{1,2}$ the two Higgs doublets. $B$ is a parameter with dimensions of mass. We  assume that ${\cal L}_{\rm dark}$ and $V$ are CP-conserving 
(i.e. $B$ and $y_\chi$ are both real, and there is no CP violation in $V_{\rm 2HDM}$) and 
we will comment on relaxing this assumption in Sec~\ref{sec:cp}. 
In this case, $a_0$ does not develop a VEV. 
We write the most general CP-conserving 2HDM potential as
\begin{align}
&V_{\rm 2HDM}=\lambda_1\left(H_1^\dagger H_1-\frac{v_1^2}{2}\right)^2+\lambda_2\left(H_2^\dagger H_2-\frac{v_2^2}{2}\right)^2
\nonumber
\\
&\quad+\lambda_3\left[\left(H_1^\dagger H_1-\frac{v_1^2}{2}\right)+\left(H_2^\dagger H_2-\frac{v_2^2}{2}\right)\right]^2
\\
&\quad+\lambda_4\left[\left(H_1^\dagger H_1\right)\left(H_2^\dagger H_2\right)-\left(H_1^\dagger H_2\right)\left(H_2^\dagger H_1\right)\right]
\nonumber
\\
&\quad+\lambda_5\left[{\rm Re}\left(H_1^\dagger H_2\right)-\frac{v_1 v_2}{2}\right]^2
+\lambda_6\left[{\rm Im}\left(H_1^\dagger H_2\right)\right]^2,
\nonumber
\end{align}
with all $\lambda_i$ real. We have also imposed a $\mathbb{Z}_2$ symmetry under which 
$H_1\to H_1$ and $H_2\to -H_2$ to suppress flavor-changing neutral currents, which 
is only softly broken by $V_{\rm 2HDM}$ and $V_{\rm port}$.  
The potential is minimized at 
$\langle H_i\rangle=v_i/\sqrt2$, $i=1,2$, and the $W$ and $Z$ masses fix 
$v_1^2+v_2^2=v^2=(246~{\rm GeV})^2$. The angle $\beta$ is defined by $\tan\beta=v_2/v_1$. 
In unitary gauge we can decompose the doublets as
\begin{align}
H_i&=\frac{1}{\sqrt{2}}
\left(\begin{array}{c}
    \sqrt 2 \phi_i^+ \\ 
    v_i+\rho_i+i\eta_i \\ 
  \end{array}\right).
\end{align}
The spectrum contains a charged Higgs,
\begin{align}
H^\pm&=\sin\beta\,\phi_1^\pm-\cos\beta\,\phi_2^\pm,
\end{align}
with mass $m_{H^\pm}^2=\lambda_4 v^2/2$. 

The CP-even Higgs mass matrix in the $(\rho_1,\rho_2)$ 
basis is ${\cal M}_h^2$, with
\begin{align}
{\cal M}_{h11}^2&=\frac{v^2}{2}\left[\lambda_5s_\beta^2 +4\left(\lambda_1+\lambda_3\right) c_\beta^2\right],
\nonumber
\\
{\cal M}_{h22}^2&=\frac{v^2}{2}\left[\lambda_5c_\beta^2 +4\left(\lambda_2+\lambda_3\right) s_\beta^2\right],
\\
{\cal M}_{h12}^2&={\cal M}_{h21}^2=\frac{v^2}{2}\left(\lambda_5+4\lambda_3\right)s_\beta c_\beta.
\nonumber
\end{align}
We use $s$ and $c$ to denote sine and cosine here (and will do so 
intermittently throughout this paper along with $t$ for tangent). The physical CP-even states 
are $h$ and $H$ $(m_h\leq m_H)$, related to $\rho_{1,2}$ by
\begin{align}
\left(\begin{array}{c}
    \rho_1 \\ 
    \rho_2 \\ 
  \end{array}\right)
&=\left(  \begin{array}{cc}
    -\sin\alpha & \cos\alpha \\ 
    \cos\alpha & \sin\alpha \\ 
  \end{array}\right)
\left(\begin{array}{c}
    h \\ 
    H \\ 
  \end{array}\right),
\\
\tan2\alpha&=\frac{2{\cal M}_{h12}^2}{{\cal M}_{h11}^2-{\cal M}_{h22}^2},
\nonumber
\end{align}
with masses
\begin{align}
m_{h,H}^2&=\frac{1}{2}\Bigg[{\cal M}_{h11}^2+{\cal M}_{h22}^2
\\
&\quad\mp\sqrt{\left({\cal M}_{h11}^2-{\cal M}_{h22}^2\right)^2+4\left({\cal M}_{h12}^2\right)^2}\Bigg].
\nonumber
\end{align}
We will use $\xi^\phi_\psi$ to denote the strength of the coupling of the scalar $\phi$ to
$\psi$ pairs (weak gauge bosons, quarks, and leptons) in units of SM Higgs coupling to 
those particles. The CP-even Higgs couplings to weak gauge bosons $V=W,Z$ are rescaled by 
\begin{align}
\xi_V^h=\sin\left(\beta-\alpha\right),~\xi_V^H=\cos\left(\beta-\alpha\right).
\end{align}

The neutral, imaginary components of $H_{1,2}$ combine to form a pseudoscalar,
\begin{align}
A_0&=\sin\beta\,\eta_1-\cos\beta\,\eta_2.
\end{align}
that mixes with $a_0$ due to the portal coupling,
\begin{align}
V_{\rm port}&=Ba_0 A_0\left[v+\sin\left(\beta-\alpha\right)h+\cos\left(\beta-\alpha\right)H\right].
\end{align}
The CP-odd mass matrix in the $(A_0,a_0)$ basis is
\begin{align}
{\cal M}_A^2&=\left(  \begin{array}{cc}
    m_{A_0}^2 & Bv \\ 
    Bv & m_{a_0}^2 \\ 
  \end{array}\right),\quad m_{A_0}^2=\frac{\lambda_6 v^2}{2}.
\end{align}
Thus, the mass eigenstates, $A$ and $a$ are
\begin{align}
&\quad\left(\begin{array}{c}
    A_0 \\ 
    a_0 \\ 
  \end{array}\right)
=\left(  \begin{array}{cc}
    \cos\theta & -\sin\theta \\ 
    \sin\theta & \cos\theta \\ 
  \end{array}\right)
\left(\begin{array}{c}
    A \\ 
    a \\ 
  \end{array}\right),
  \nonumber
\\
&\quad\tan2\theta=\frac{2Bv}{m_{A_0}^2-m_{a_0}^2},
\\
m_{a,A}^2&=\frac12\left[m_{A_0}^2+m_{a_0}^2
\pm\sqrt{\left(m_{A_0}^2-m_{a_0}^2\right)^2+4B^2v^2}\right].
\nonumber
\end{align}
We can express $B$ in terms of $m_{a,A}$ and the mixing angle $\theta$,
\begin{align}
B&=\frac{1}{2v}\left(m_{A}^2-m_{a}^2\right)\sin2\theta.
\end{align}

Written in terms of mass eigenstates and mixing angles, $V_{\rm port}$ becomes
\begin{align}
V_{\rm port}&=\frac{1}{2v}\left(m_{A}^2-m_{a}^2\right)\left[s_{4\theta}aA+s_{2\theta}^2\left(A^2-a^2\right)\right]
\nonumber
\\
&\times\left[\sin\left(\beta-\alpha\right)h+\cos\left(\beta-\alpha\right)H\right].
\label{eq:vport-mass}
\end{align}

The DM coupling to the mediator in Eq.~(\ref{eq:dm-yuk-flavor}) is simply expressed in terms of 
CP-odd mass eigenstates,
\begin{align}
{\cal L}_{\rm dark}&=y_\chi \left(\cos\theta\,a+\sin\theta\,A\right)\bar\chi i\gamma^5\chi.
\label{eq:dm-yuk-mass}
\end{align}

We will work in a Type II 2HDM, where the Yukawa couplings of the SM fermions are
\begin{align}
{\cal L}_{\rm Yuk}&=-\bar L Y_e H_1 e_R-\bar Q Y_d H_1 d_R-\bar Q Y_d \tilde H_2 u_R+{\rm h.c.}.
\nonumber
\end{align}
$Y_{e,d,u}$ are Yukawa matrices acting on the three generations (we employ first generation 
notation). $L$ and $Q$ are the left-handed lepton and quark doublets and $e_R$, $d_R$, and $u_R$ are the
right-handed singlets.  These couplings respect the $\mathbb Z_2$ symmetry $H_2\rightarrow -H_2$ provided $u_R\rightarrow -u_R$.  We can forbid the operator
\begin{align}
{\cal L}_{\rm Yuk}&=-\bar L Y_\chi \tilde{H}_1\chi_R  +{\rm h.c.}.
\nonumber
\end{align} by taking $\chi\rightarrow-\chi$ under a separate $\mathbb{Z}_2$.
Note $\tilde H_i$ stands for $i\sigma_2H_i^\ast$. Given these Yukawa 
interactions the couplings of the neutral 
scalar mass eigenstates are then rescaled from the SM Higgs values by
\begin{align}
\xi_e^h&=\xi_d^h=-\frac{\sin\alpha}{\cos\beta},~\xi_u^h=\frac{\cos\alpha}{\sin\beta},
\\
\xi_e^H&=\xi_d^H=\frac{\cos\alpha}{\cos\beta},~\xi_u^H=\frac{\sin\alpha}{\sin\beta},
\\
\xi_e^A&=\xi_d^A=\tan\beta\cos\theta,~\xi_u^A=\cot\beta\cos\theta,
\\
\xi_e^a&=\xi_d^a=-\tan\beta\sin\theta,~\xi_u^a=-\cot\beta\sin\theta.
\end{align}

To simplify the analysis, we work close to the decoupling limit of the 2HDM where 
\begin{align}
\lambda_1\simeq\lambda_2\simeq-\lambda_3\simeq\frac{\lambda_4}{2}\simeq\frac{\lambda_5}{2}\simeq\frac{\lambda_6}{2}\equiv\lambda\gg1.
\end{align}
Then, $\alpha\simeq\beta-\pi/2$ and $m_h\ll m_H\simeq m_{H^\pm}\simeq m_{A_0}$. 
Since $h$ has SM-like couplings in this limit, we identify it with the boson with mass 125 GeV recently 
discovered at the LHC. The degeneracy of $H$ and $H^\pm$ (and possibly $A$, given that we expect $B$ to be somewhat small compared to $m_{A_0}$) allows for precision electroweak constraints to be satisfied.

\subsection{Dark Matter CP Problem}
\label{sec:cp}
We now briefly discuss relaxing the assumption of CP conservation in the 
DM Yukawa interaction or in the 
scalar potential.
If we write a general, possibly CP-violating, 4-Fermi interaction 
between quarks and DM 
that results after integrating out a spin-0 mediator as
\begin{align}
{\cal L}_{\rm eff}=\frac{1}{\Lambda^2}\frac{m_q}{v}\bar\chi
\left(a_\chi+ib_\chi\gamma^5\right)\chi\bar q\left(a_q+ib_q\gamma^5\right)q,
\label{eq:dim6}
\end{align}
we find an annihilation cross section for $\chi\bar\chi\to b\bar b$ in the 
nonrelativistic limit, relevant for thermal freeze-out and indirect detection, of
\begin{align}
\sigma v_{\rm rel}&=\frac{1}{2\pi}\left(\frac{m_\chi m_b}{\Lambda^2v}\right)^2
\left(b_\chi^2+a_\chi^2v_{\rm rel}^2\right)\left(b_b^2+a_b^2\right)
\nonumber
\\
&\simeq3\times10^{-26}~\frac{\rm cm^3}{\rm s}\left(\frac{m_\chi}{30~\rm GeV}\right)^2\left(\frac{54~\rm GeV}{\Lambda}\right)^4
\\
&\quad\times\left(b_\chi^2+a_\chi^2v_{\rm rel}^2\right)\left(b_b^2+a_b^2\right),
\nonumber
\end{align}
with $v_{\rm rel}$ the relative velocity between $\chi$ and $\bar\chi$. We have taken $m_b\ll m_\chi$ and normalized on parameters that give the 
appropriate annihilation cross section for a thermal relic as well as the gamma ray 
excess. This operator also leads to a 
spin-independent cross section for DM scattering on a nucleon of
\begin{align}
\sigma_{\rm SI}&=\frac{\mu^2}{\pi}\left(\frac{\bra{N}\sum_q a_q m_q\bar qq\ket{N}}{\Lambda^2v}\right)^2a_\chi^2
\nonumber
\\
&\simeq 2.6\times10^{-41}~{\rm cm}^2\left(\frac{54~\rm GeV}{\Lambda}\right)^4
\\
&\quad\times\left(\frac{\bra{N}\sum_q a_q m_q\bar qq\ket{N}}{330~\rm MeV}\right)^2a_\chi^2,
\nonumber
\end{align}
where $\mu$ is the reduced mass of the nucleon-DM system. The LUX experiment 
has set a limit of $\sigma_{\rm SI}<8\times10^{-46}~{\rm cm}^2$~\cite{Akerib:2013tjd} 
at a dark matter mass of $30~{\rm GeV}$, which highlights a problem for the general
dimension-six operator in Eq.~(\ref{eq:dim6}). The scalar-scalar coupling needs to be  
suppressed by about five orders of magnitude relative to the 
pseudoscalar-pseudoscalar coupling (which is why the latter 
has been focused on in the literature) without any good reason. If not, the 
stringent limits 
from direct detection preclude the possibility of an annihilation cross 
section large enough for an observable indirect 
detection signal or even to obtain the observed relic density and not overclose 
the Universe. 

A scalar coupling of $a$ to $\chi$ is obtained if $y_\chi$ in 
Eq.~(\ref{eq:dm-yuk-flavor}) has a nonzero imaginary component. $a$ will also develop 
a scalar coupling to quarks if $B$ in Eq.~(\ref{eq:Vport}) is not real or if there 
is CP violation in the rest of the scalar potential. As highlighted above in the 
discussion of a general dimension-six interaction, (the product of) these 
CP-violating phases in $y_\chi$ and $V$ must be limited to less than about 
$10^{-4}$ to $10^{-5}$ (ignoring possible suppressions or enhancements at large 
$\tan\beta$). That is, in addition to the usual EDM constraints on CP-violating 
phases in the scalar potential (see, e.g.~\cite{McKeen:2012av,*Ipek:2013iba}), these 
models face tests from direct detection experiments (which become probes of 
CP violation).

CP is not an exact symmetry of the SM; indeed, we expect to see 
even larger violations of CP in physics beyond the SM because of 
baryogenesis. In this light, simply asserting that these new interactions respect 
CP seems a little peculiar. It is however technically natural to 
assume that spurions proportional to the SM Yukawa coupling matrices 
are the only source of CP and flavor violation, 
with the consequence that CP-violating couplings outside of the CKM are tiny.

Should the evidence for this gamma ray signal remain or increase, understanding 
the smallness of these CP-violating couplings will pose a model-building 
challenge and hint about the structure of the new physics, 
much like the CP problems encountered in other models of physics beyond the SM.

\section{Dark Matter Annihilation}
\label{sec:ann}
For $m_a\ll m_A$, the dark matter annihilates to SM particles primarily through $s$-channel $a$ exchange. The velocity averaged annihilation cross section for $\chi\bar\chi\to{\rm SM}$ in 
the nonrelativistic limit is
\begin{align}
\langle \sigma v_{\rm rel}\rangle&=\frac{y_\chi^2}{8\pi}\frac{m_\chi^2}{m_a^4}s_{2\theta}^2 \tan^2\beta\left[\left(1-\frac{4m_\chi^2}{m_a^2}\right)^2+\frac{\Gamma_a^2}{m_a^2}\right]^{-1}
\nonumber
\\
&\quad\quad\quad\quad\times\sum_{f=b,\tau,\dots} N_C \frac{m_f^2}{v^2}\sqrt{1-\frac{m_f^2}{m_a^2}}.
\label{eq:annih}
\end{align}
The sum is over down-type quarks ($N_C=3$) and charged leptons ($N_C=1$) since $a$'s coupling to up-type quarks is suppressed by $1/\tan\beta$. Evaluating this 
at the experimentally favored DM mass of 30~GeV, taking $m_a=100~{\rm GeV}$ (and ignoring
$\Gamma_a$) gives
\begin{align}
\langle \sigma v_{\rm rel}\rangle&=3\times10^{-26}~\frac{\rm cm^3}{\rm s}\left(\frac{y_\chi \sin2\theta \tan\beta}{2.4}\right)^2.
\end{align}
We see that it is possible to achieve values of the annihilation cross section 
compatible with the gamma ray excess and relic density constraints with modest values of 
the mixing angle $\theta$, provided $\tan\beta$ is somewhat large. At this value of 
$m_\chi$ and for $\tan\beta$ larger than a few, the $b\bar b$ final 
state accounts for about 90\% of the annihilation cross section with $\tau^+\tau^-$ 
making up nearly all the rest. This is in line with what is suggested by fits to the gamma ray excess.

The general requirement that $\tan\beta$ is large will help focus the mass scale of the 
heavy Higgs bosons. CMS's search for heavy neutral minimal supersymmetric standard 
model (MSSM) Higgses decaying to 
$\tau^+\tau^-$~\cite{CMS-PAS-HIG-13-021} applies straightforwardly in this case, since
the Higgs sector is the same as in the MSSM. The production cross section for 
$pp\to (H/A)+X$ is enhanced at large $\tan\beta$ so the lack of a 
signal sets an upper limit on $\tan\beta$ as a function of $m_{A,H}$. 
This limit is roughly $\tan\beta<10$ at $m_{A,H}=300~{\rm GeV}$, and weakens to 
$\tan\beta<60$ at $m_{A,H}=900~{\rm GeV}$.

\section{Constraints on the Dark Sector}
\label{sec:results}
\begin{figure}[t!]
\includegraphics[scale=0.6]{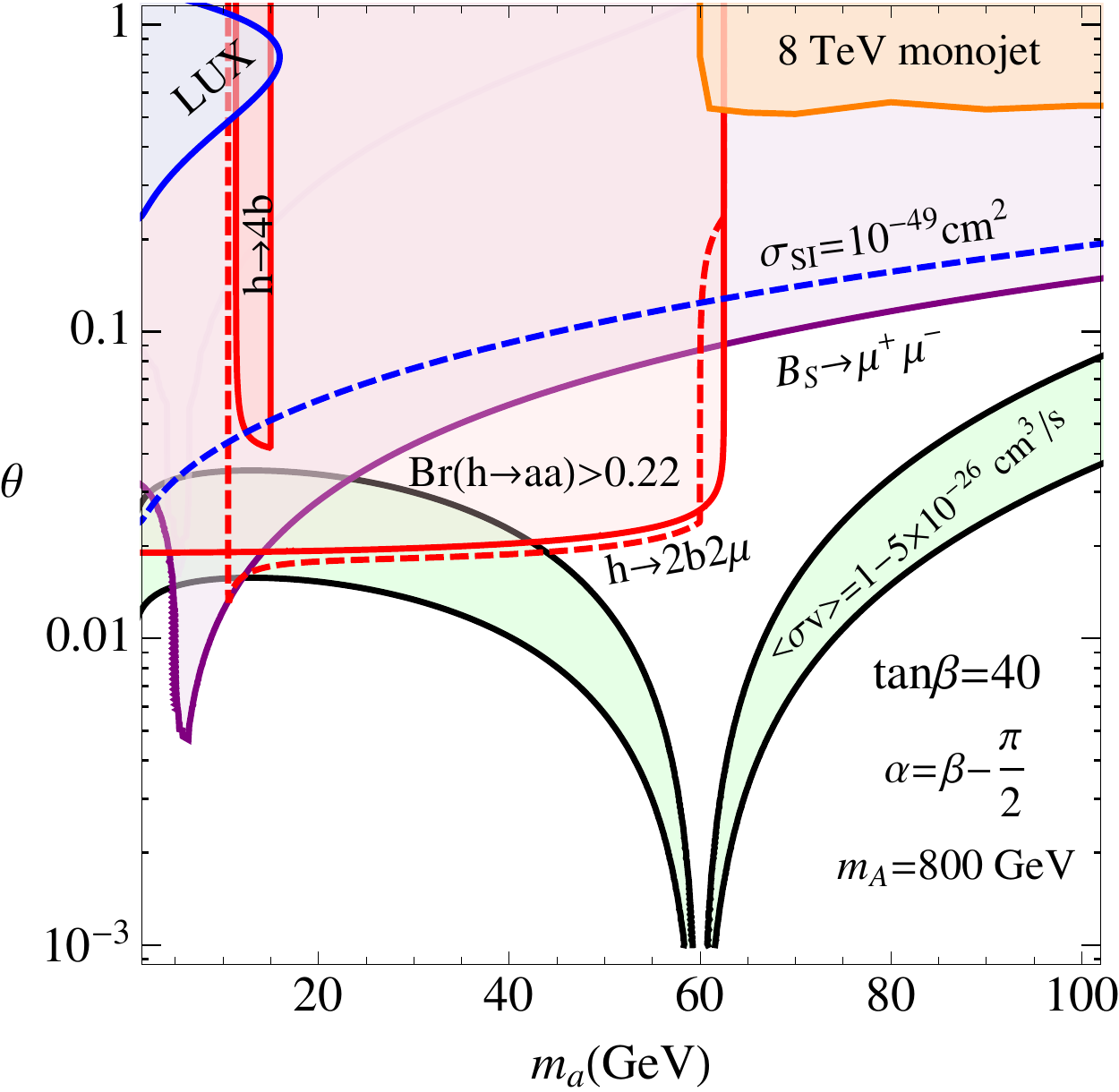}
\caption{Regions of mixing angle $\theta$ vs. $m_a$ that are ruled out or 
suggested by various measurements. We have fixed 
$m_{H,H^\pm}\simeq m_A=800~{\rm GeV}$, $\tan\beta=40$, $\alpha=\beta-\pi/2$, 
and $y_\chi=0.5$.
The area that gives an 
annihilation cross section
of $\langle \sigma v_{\rm rel}\rangle=1-5\times10^{-26}~{\rm cm}^3/{\rm s}$ 
as indicated by fits to the gamma ray excess is between the solid black lines 
(shaded in green).  
The shaded purple region above the solid purple line is in $2\sigma$ conflict with the LHCb measurement of 
$B_s\to\mu^+\mu^-$. The darker red region 
with the solid outline is ruled out by $h\to b\bar b$ constraints on the 
$h\to 4b$ signal. The larger, lighter red region with a solid outline is ruled out from the indirect limit $\Br\left({h\to aa}\right)<0.22$ coming from fits to Higgs properties, assuming SM Higgs production. The dashed red line shows the area that could be probed by 
limiting $\Br\left({h\to aa\to 2b2\mu}\right)\lesssim 10^{-4}$. The blue region 
labeled LUX is in conflict with the limit 
$\sigma_{\rm SI}<8\times10^{-46}~{\rm cm}^2$ while the area above the blue 
dashed line leads to $\sigma_{\rm SI}>10^{-49}~{\rm cm}^2$, potentially accessible 
at the next generation of direct detection experiments. The orange region shows the 
area ruled out by a mono-$b$-jet search at 8 TeV with 20~fb$^{-1}$ of data. See text for details.
\label{fig:excl}}
\end{figure}

In this section we investigate the limits on the mediator mass and the mixing 
angle between the mediator and the pseudoscalar of the 2HDM. Taking the heavy Higgs search described above into account, we fix the other 
parameters to the benchmark values $m_H=m_{H^\pm}\simeq m_A=800~{\rm GeV}$, $\tan\beta=40$, $\alpha=\beta-\pi/2$, 
and $y_\chi=0.5$ and comment on changing these later. We first consider the spin-independent direct detection cross section induced at one-loop. Current limits from direct detection experiments do not constrain this model, but future searches can possibly probe interesting regions of parameter space. We next consider Higgs decays to the pseudoscalar mediator. Searches for $h\to b\bar{b}$ can be used to put bounds to $h\to aa \to 4b$ decays for $m_h>2m_a$ and future $h\to 2b 2\mu$ searches could probe much more of the $m_a$-$\theta$ parameter space. Indirect limits on the branching for $h\to aa$ from global Higgs property fits are also quite constraining. We then consider changes to the $B_s\to\mu^+\mu^-$ branching ratio. Since this has been measured to be very close to its SM value, it is particularly constraining for a light mediator. Finally, we consider monojet searches. Our main results are summarized in Fig.~\ref{fig:excl}. 

\subsection{Direct Detection}
\label{sec:dd}
\begin{figure}
\includegraphics[scale=0.6]{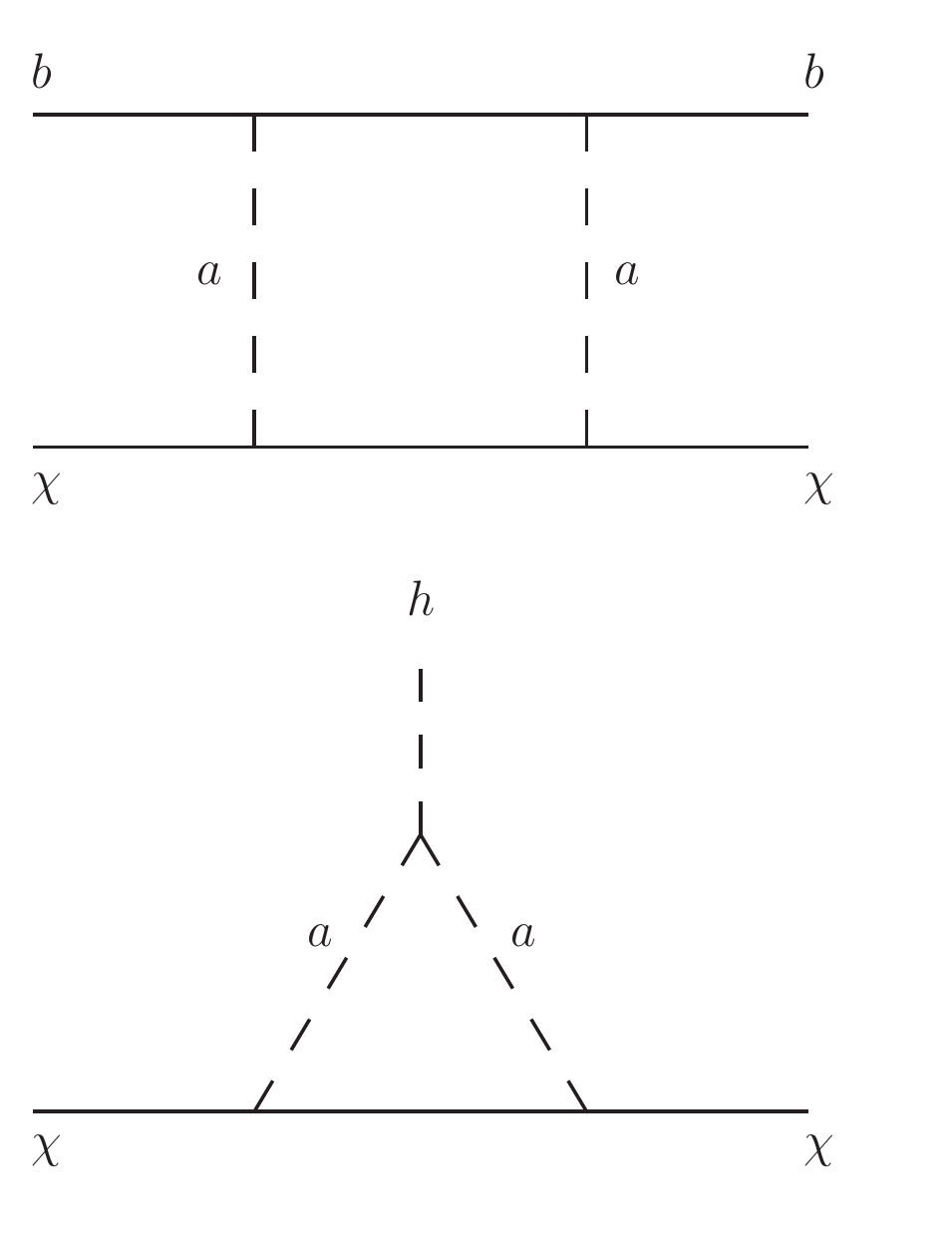}
\caption{Diagrams that (with the crossed top diagram) lead to spin-independent 
DM-nucleon scattering. The $b$ quarks in the top diagram can also be 
replaced by $d$ and $s$ quarks.
\label{fig:1loop}}
\end{figure}

One of the virtues of this model is that 
single pseudoscalar exchange between $\chi$ and quarks leads to (highly suppressed) spin-dependent scattering 
of the DM on nuclei~\cite{Boehm:2014hva,Alves:2014yha}. At one-loop, 
however, spin-independent interactions are generated through the diagrams shown in 
Fig.~\ref{fig:1loop}.
The top diagram (plus its crossed version) leads to an effective interaction between 
$\chi$ and $b$ quarks at zero momentum transfer given by
\begin{align}
{\cal L}_{\rm box}&=\sum_{q=d,s,b}\frac{m_q^2 y_\chi^2\tan^2\beta \sin^22\theta}{128\pi^2m_a^2\left(m_\chi^2-m_q^2\right)}
\label{eq:Lbox}
\\
&\quad\quad\times\left[F\left(\frac{m_\chi^2}{m_a^2}\right)-F\left(\frac{m_q^2}{m_a^2}\right)\right]\frac{m_\chi m_q}{v^2}\bar\chi \chi \bar q q.
\nonumber
\end{align}
The function $F$ is given in the Appendix in Eq.~\ref{eq:F}.

The bottom diagram of Fig.~\ref{fig:1loop} leads to a DM-Higgs coupling of
\begin{align}
{\cal L}_{h\chi\chi}&=-\frac{\left(m_A^2-m_a^2\right)\sin^2{2\theta} y_\chi^2}{64\pi^2 m_a^2}G\left(x_\chi,x_q\right)\frac{m_\chi}{v}h\bar\chi\chi,
\label{eq:Lhchichi}
\end{align}
where $x_\chi=m_\chi^2/m_a^2$, $x_q=q^2/m_a^2$, and $q$ is the momentum 
transfer between $\chi$ and $\bar\chi$. $G$ is given in Eq.~\ref{eq:G}. This leads to 
an effective 4-fermion interaction relevant for spin-independent nucleon 
scattering,
\begin{align}
{\cal L}_{h}&=\frac{\left(m_A^2-m_a^2\right)s_{2\theta}^2 y_\chi^2}{64\pi^2 m_h^2m_a^2}G\left(x_\chi,0\right)\frac{m_\chi m_q}{v^2}\bar\chi\chi \bar qq.
\end{align}
We have assumed $\alpha=\beta-\pi/2$ which results in SM-like couplings of $h$ 
to quarks, $-s_\alpha/c_\beta=c_\alpha/s_\beta=1$.
For
\begin{align}
\tan\beta\lesssim 100\left(\frac{m_A}{800~{\rm GeV}}\right),
\end{align}
the Higgs exchange contribution to direct detection dominates over the box 
diagram, leading to a  
spin-independent cross section for scattering on a nucleon of
\begin{align}
&\sigma_{\rm SI}\simeq 2.2\times10^{-49}~{\rm cm^2}\left(\frac{m_A}{800~{\rm GeV}}\right)^4\left(\frac{50~{\rm GeV}}{m_a}\right)^4
\\
&\times\left(\frac{m_\chi}{30~{\rm GeV}}\right)^2\left(\frac{\theta}{0.1}\right)^4\left(\frac{y_\chi}{0.5}\right)^4\left(\frac{\bra{N}\sum_qm_q\bar qq\ket{N}}{330~{\rm MeV}}\right)^2,
\nonumber
\end{align}
using a value of the 
$\bar qq$ matrix element from Ref.~\cite{Ellis:2000ds}. Cross sections at the 
$10^{-48}$ to $10^{-49}~{\rm cm^2}$ level are potentially 
observable at the next generation of direct detection 
experiments~\cite{Malling:2011va}. The loop suppression of the spin-independent cross 
section, however, is sufficient for this model to remain safe from direct detection 
experiments for the near future.

In Fig.~\ref{fig:excl}, we show the area of parameter space ruled out by the LUX 
limit of $8\times10^{-46}~{\rm cm^2}$ on a spin-independent cross section that 
arises from the loop diagrams above. Only areas of very large mixing angle 
$\theta$ and small $m_a$ are impacted. We also show the area that can be probed 
by a future cross section limit of $10^{-49}~{\rm cm^2}$ which covers a much larger region.

\subsection{Higgs Decays}
\label{sec:higgs}
If $m_a<m_h$, the Higgs can decay into final states involving $a$ and, in 
particular, when $m_a<m_h/2$, the two body mode $h\to aa$ opens up. Using  Eq.~(\ref{eq:vport-mass}) with $\sin\left(\beta-\alpha\right)=1$, the rate is
\begin{align}
\Gamma\left({h\to aa}\right)&=\frac{\left(m_A^2-m_a^2\right)^2\sin^4 2\theta}{32\pi m_h v^2}\sqrt{1-\frac{4m_a^2}{m_h^2}}
\\
&\simeq840~{\rm MeV}\left(\frac{m_A}{800~{\rm GeV}}\right)^4\left(\frac{\theta}{0.1}\right)^4,
\nonumber
\end{align}
having taken $m_a\ll m_h,\,m_A$ in the 
second line. Since the width of the SM Higgs is 4~MeV, this can impact LHC measurements 
that broadly indicate that $h$ is SM-like to $\sim$10-20\% if $\theta\gtrsim{\rm few}\times10^{-2}(800~{\rm GeV}/m_A)$.

This mode requires $m_a<m_h/2\simeq 2m_\chi$, so the pseudoscalars will primarily go to $b$ quarks with a small branching to $\tau$ and $\mu$ pairs. The $h\to aa\to4b$ signal will contribute to $h\to b\bar b$ searches~\cite{Curtin:2013fra}. A CMS search in $W/Z$-associated production at 7 and 8~TeV, 
$pp\to (W/Z)+(h\to b\bar b)$~\cite{Chatrchyan:2013zna}, sets a limit
$\Br\left({h\to aa\to4b}\right)<0.7$ for $2m_b<m_a<15~{\rm GeV}$. 
This can potentially be improved to 
$0.2$ with 100~fb$^{-1}$ data at the 14~TeV LHC. For larger $m_a$, 
there are no current limits. Additionally, the $h\to 2b2\mu$ final state can offer a 
probe comparable to $4b$, with its relative cleanliness compensating for its rarity. Current 7 and 8~TeV data can 
limit this to $\Br\left({h\to aa\to 2b2\mu}\right)\lesssim 10^{-3}$ for $m_a>25~{\rm GeV}$. The 14 TeV run with 
100~fb$^{-1}$ data can improve this limit to $10^{-4}$~\cite{Curtin:2013fra}.

Taking the branching ratios of $a$ into account, the limits above translate into a 
limit $\Br\left(h\to aa\right)\lesssim 0.9$ for $2m_b<m_a<15~{\rm GeV}$ currently, 
with the possibility of improving this to $\Br\left(h\to aa\right)\lesssim 0.1-0.2$ in 
the future. 

Since we are in the decoupling limit, the production cross section of the Higgs is unchanged from its SM value in this model (unless we add further states). Therefore there are strong limits on unobserved final states, such as $aa$, that would dilute the signal strength in the observed channels. Given current data, this limits $\Br\left(h\to aa\right)<0.22$~\cite{Giardino:2013bma}.

The decay $h\to\chi\bar\chi$ through the bottom diagram in Fig.~\ref{fig:1loop} is 
loop-suppressed and offers no meaningful constraints. For larger 
$m_a$, the three-body decays $h\to aa^\ast,\,aA^\ast$ become the dominant exotic 
Higgs decay modes, but are suppressed by the extra particle in the final state and 
are also not constraining.

We show the limits on the mixing angle as a function of $m_a$ coming from the 
determination $\Br\left({h\to aa\to4b}\right)<0.7$ as well as the indirect constraint $\Br\left(h\to aa\right)<0.22$ in Fig.~\ref{fig:excl}. We also 
show the limit that can be set by a future measurement of 
$\Br\left({h\to aa\to 2b2\mu}\right)<10^{-4}$. $h\to aa$ decays provide strong 
constraints when kinematically allowed, i.e. $m_a\lesssim 60~{\rm GeV}$. 

\subsection{B Physics Constraints}
\label{sec:bphys}
A light $a$ can also potentially be constrained by its contributions to the decay 
$B_s\to\mu^+\mu^-$. For $m_a\ll m_Z$, the correction 
due to s-channel $a$ exchange can be simply written as~\cite{Skiba:1992mg}
\begin{align}
&\Br\left(B_s\to\mu^+\mu^-\right)\simeq\Br\left(B_s\to\mu^+\mu^-\right)_{\rm SM}
\\
&\quad\quad\quad\quad\times\left|1+\frac{m_b m_{B_s}t_\beta^2 s_\theta^2}{m_{B_s}^2-m_a^2}\frac{f\left(x_t,y_t,r\right)}{Y\left(x_t\right)}\right|^2,
\nonumber
\end{align}
with $x_t=m_t^2/m_W^2$, $y_t=m_t^2/m_{H^\pm}^2$, $r=m_{H^\pm}^2/m_W^2$,
\begin{align}
f\left(x,y,r\right)&=\frac{x}{8}\Bigg[-\frac{r\left(x-1\right)-x}{\left(r-1\right)\left(x-1\right)}\log r+\frac{x \log x}{\left(x-1\right)}
\nonumber
\\
&\quad-\frac{y \log y}{\left(y-1\right)}+\frac{x \log y}{\left(r-x\right)\left(x-1\right)}\Bigg],
\end{align}
and $Y(x)$ the usual Inami-Lim function,
\begin{align}
Y\left(x\right)&=\frac{x}{8}\left[\frac{x-4}{x-1}\log x +\frac{3x \log x}{\left(x-1\right)^2}\right].
\end{align}
The average of the LHCb and CMS measurements of this mode is
$\Br\left(B_s\to\mu^+\mu^-\right)=\left(2.9\pm0.7\right)\times10^{-9}$~\cite{Aaij:2013aka,*Chatrchyan:2013bka,*CMS-PAS-BPH-13-007}. 
This should be compared against the SM prediction, which we take to be
$\left(3.65\pm0.23\right)\times10^{-9}$~\cite{Bobeth:2013uxa,*Buras:2013uqa}. This offers a strong 
test of the model, especially for a light $a$, which we show in Fig.~\ref{fig:excl}.

\subsection{Collider Probes}
\label{sec:colliders}
Monojet and monophoton searches have become standard techniques to look for 
dark matter at hadron colliders in recent years (see, e.g.,~\cite{Beltran:2010ww,*Goodman:2010ku,*Rajaraman:2011wf,*Fox:2011pm,*Shoemaker:2011vi,*Fox:2012ee,*Chatrchyan:2012me,*ATLAS:2012ky}).

To estimate the reach that such searches might have in this model, we make 
use of a recent analysis~\cite{Lin:2013sca} designed to probe dark 
matter that 
couples preferentially to heavy quarks, taking advantage of $b$-tagging 
a jet recoiling against missing energy to cut down substantially on 
backgrounds.

For our signal, we use MadGraph 5~\cite{Alwall:2011uj} to generate matched samples of $\chi\bar\chi+(0,1,2)j$ (with $j$ representing 
both $b$- and light flavor/gluon-jets), 
shower and hadronize with \textsc{Pythia} 6~\cite{Sjostrand:2006za}, 
and use \textsc{Delphes} 2~\cite{Ovyn:2009tx} for detector simulation. We 
take a 50\% $b$-tag efficiency for $p_T>80~{\rm GeV}$, which measurements 
from CMS~\cite{Chatrchyan:2012jua} and ATLAS~\cite{ATLAS-CONF-2014-004} show is conservative (a larger efficiency would lead to stronger limits).

The most useful signal region defined in Ref.~\cite{Lin:2013sca} 
for this model has the following requirements: (i) missing transverse energy 
greater than 350~GeV, (ii) no more than two jets with $p_T>50~{\rm GeV}$, 
(iii) the leading jet has $p_T>100~{\rm GeV}$ and is $b$-tagged, (iv) no 
isolated leptons, and (v) if there is a second jet, its separation in 
azimuthal angle with respect to the missing energy is $\Delta\phi>0.4$.
Using backgrounds estimated in~\cite{Lin:2013sca} at the 8 TeV LHC
(mainly $Z$+jets and $t\bar t$+jets), 
we can identify regions of parameter space that can be expected to be probed by 
this search with 20~fb$^{-1}$ of data. Monojet searches of this type are 
most sensitive when 
$m_a>2m_\chi$ since then $a$ can be produced on-shell and decay to 
$\chi\bar\chi$. If $m_a<2m_\chi$ the reach substantially weakens due to the 
additional particle in the final state and the softness of 
the missing energy since the $\chi\bar\chi$ pair tends to be created close 
to threshold.

This model is relatively less well constrained by monojet searches than 
the EFTs studied in Refs.~\cite{Boehm:2014hva} and~\cite{Lin:2013sca} 
because of the suppressed coupling to top at large $\tan\beta$.
The values of $\theta$ as a function of $m_a$ that would be ruled out by the 
search described above 
are shown in Fig.~\ref{fig:excl}. Extending this search to 100~fb$^{-1}$ of 
14~TeV data could improve the reach in $\theta$ by a factor of 
$\sim 3$~\cite{Lin:2013sca}.

\section{Conclusions and Outlook}
\label{sec:conc}
An excess in gamma rays from the Galactic Center as measured by the Fermi Gamma Ray Space Telescope can be explained by 30~GeV DM annihilating dominantly into $b\bar{b}$ pairs. To do so while eluding bounds on spin-independent scattering of DM on nuclei suggests that the mediator between the dark sector and the SM is a pseudoscalar. We have studied a 2HDM where the pseudoscalar mediator mixes with the CP-odd Higgs, giving rise to interactions between DM and the SM. 

At one-loop, scalar-scalar interactions between DM and SM quarks arise. This leads to a spin-independent cross section for direct detection well below the current bound of $8\times10^{-46}{\rm cm}^2$ at a dark matter mass of 30~GeV. Future limits at better than $10^{-49}{\rm cm}^2$ could impact this model. We also consider decays of the 125~GeV SM-like Higgs boson involving the mediator. If the mediator is light $h\to aa\to 4b,\,2b2\mu$ can be constraining with data from the 14~TeV LHC. Additional contributions to $B_s\to\mu^+\mu^-$ in this model eliminate some of the favored parameter space for $m_a<10~{\rm GeV}$. This scenario is not well tested by monojet searches, including ones that rely on $b$-tagging to increase the sensitivity to DM coupled to heavy quarks, due to a suppressed coupling of the mediator to $t$ quarks. 

Changing the benchmark parameters that we used above does not greatly change 
the general results. For example, if we lower lower $m_A$ to decrease the 
$h\to a a$ signal
coming from Eq.~(\ref{eq:vport-mass}), we have to decrease $\tan\beta$ because 
of the CMS heavy Higgs search~\cite{CMS-PAS-HIG-13-021}. Then, to obtain 
the correct annihilation cross section in Eq.~(\ref{eq:annih}), we have to increase 
the mixing angle (or, equivalently, $B$) which in turn increases the $h\to a a$ rate.

One obvious piece of evidence in favor of this scenario would be finding heavy 
Higgses at the LHC. However, conclusively determining whether these heavy Higgses 
are connected to 30~GeV DM annihilating at the center of the galaxy will be a 
formidable challenge. Among the possible signatures to probe this scenario is 
$A\to h a\to 2b+{\rm inv.}$ We leave a detailed study of this search and others 
for future work.

\begin{acknowledgements}
This work was supported in part by the U.S. Department of Energy under 
Grant No. DE-FG02-96ER40956. We thank David Morrissey for useful discussions 
during the early stages of this work.

\end{acknowledgements}

\textit{Note added}.---While this work was in preparation Refs.~\cite{Izaguirre:2014vva,Fedderke:2014wda} appeared, with some overlapping results.

\appendix*
\section{Loop Functions}
We provide expressions for the loop functions presented above in this appendix.

The form factor needed for the box diagrams in Eq.~(\ref{eq:Lbox}) is given by 
\begin{align}
F\left(x\right)&=\frac{2}{3x}\left[4+f_+\left(x\right)+f_-\left(x\right)\right],
\label{eq:F}
\end{align}
with
\begin{align}
f_\pm\left(x\right)&\equiv \frac{1}{x}\left(1\pm\frac{3}{\sqrt{1-4x}}\right)\left(\frac{1\pm\sqrt{1-4x}}{2}\right)^3
\\
&\quad\quad\times\log\left(\frac{1\pm\sqrt{1-4x}}{2}\right).
\nonumber
\end{align}
$F$ vanishes when its argument is small as $F(x\to 0)\to(4x/3)\log(1/x)$.

The function arising in the effective $\chi$-$\chi$-$h$ interaction in 
Eq.~(\ref{eq:Lhchichi}) is
\begin{align}
G\left(x,y\right)&=-4i\int_0^1dz\,\frac{z}{\lambda^{1/2}\left(x,y,z\right)}
\label{eq:G}
\\
&\quad\times\log\left[\frac{\lambda^{1/2}\left(x,y,z\right)+iy\left(1-z\right)}{\lambda^{1/2}\left(x,y,z\right)-iy\left(1-z\right)}\right],
\nonumber
\end{align}
with
\begin{align}
\lambda\left(x,y,z\right)&\equiv y\left[4\left(1-z\right)+4xz^2-y\left(1-z\right)^2\right].
\end{align}
For small arguments $G$ approaches unity, $G\left(0,0\right)=1$.
\bibliography{ref}

\end{document}